\documentclass{aa}  

\usepackage{natbib}
\bibpunct{(}{)}{;}{a}{}{,} 
\usepackage{graphicx}
\usepackage{txfonts}
\usepackage{txfonts}
\usepackage{color}
\usepackage{colortbl}
\usepackage{placeins}
\usepackage{float}
\usepackage{subfigure}

\usepackage{xcolor}

\usepackage{arydshln}      
\begin{document} 

   \title{Studies of stationary features in jets: 3C~279 quasar I. On-sky scattering and dynamics}

   \authorrunning{Arshakian et al.}
   \titlerunning{Studies of stationary features in jets: 3C 279 quasar}
   \subtitle{}

   \author{T.G. Arshakian
          \inst{1,2,3},
          L.A. Hambardzumyan
          \inst{2,3}, 
          A.B. Pushkarev
          \inst{4,5},
          D.C. Homan
          \inst{6},
          E.L. Karapetyan 
          \inst{2}
          }

   \institute{I. Physikalisches Institut, Universität zu Köln, Zülpicher Strasse 77, Köln, Germany, \\
              \email{arshakian@ph1.uni-koeln.de}
         \and
            Astrophysical Research Laboratory of Physics Institute, Yerevan State University, 1 Alek Manukyan St., Yerevan, Armenia         
        \and
            Byurakan Astrophysical Observatory after V.A. Ambartsumian, Aragatsotn Province 378433, Armenia
        \and
            Crimean Astrophysical Observatory, 298409 Nauchny, Crimea
        \and 
            Lebedev Physical Institute of the Russian Academy of Sciences, Leninsky prospekt 53, 119991 Moscow, Russia
        \and
            Department of Physics and Astronomy, Denison University, Granville, OH 43023, USA
             }

   \date{Received xxx; accepted xxx}

 
  \abstract
  {A recent study on the dynamics of the quasi-stationary component (QSC) in the jet of BL~Lacertae highlighted its significance in evaluating the physical properties of relativistic transverse waves in the parsec-scale jet. Motivated by this finding, we selected a different type of blazar, the flat-spectrum radio quasar (FSRQ) 3C~279, which hosts a QSC at an angular median distance of 0.35~mas from the radio core, as has been revealed by 27 years of VLBA monitoring data at 15~GHz.}
  {We investigate the positional scatter and dynamics of a QSC in the 3C~279 jet, aiming to detect the presence of a relativistic transverse wave and estimate its characteristics.}
  {We employed an analytical statistical method to estimate the mean intrinsic speed of the QSC, while moving average and refinement methods were used to smooth its trajectory. }
  {Analysis of the QSC position scatter shows that the jet axis changes its direction by about $21\degr$ over 27 years and jet mean intrinsic full opening angle is $\approx 0.30\degr \pm 0.03\degr$. The apparent displacement vectors of the QSC exhibit strong asymmetry and anisotropy in the direction of the jet, indicating pronounced anisotropic displacements of the core along the jet axis. We estimated the mean intrinsic speed of the QSC to be superluminal, $\overline{\beta_{\rm s}} \approx 10$ in units of the speed of light, which, within the framework of the seagull-on-wave model, is interpreted as evidence of a relativistic transverse wave propagating through the QSC. Analysis of the reversing trajectory of the QSC enables the classification and characterization of reversal patterns, which, in turn, allows us to determine key transverse wave parameters such as the frequency, amplitude, inclination angle, and magnetic energy of the wave. }
  {Analysing the variations in the QSC position and dynamics provides a powerful diagnostic for determining changes in the jet cone angle and direction, for detecting relativistic transverse waves in the jets of two types of blazars, BL Lacertae and the FSRQ 3C 279, and for characterising the properties of these transverse waves.
  
  }  

   \maketitle
%

\section{Introduction}
\label{sec:int}
The proximity of the BL~Lacertae object (hereafter, BL~Lac) and its relatively high radio brightness have motivated a detailed study of the structure and kinematics of its relativistic jet using radio interferometric observations. \cite{cohen14,cohen15} conducted a comprehensive analysis of the BL~Lac jet using VLBA monitoring data at 15~GHz over 17 years. They identified a quasi-stationary component (QSC) located at a projected distance of $\sim 0.26$~mas from the core, along with moving components emanating downstream of the QSC. An examination of the jet ridge line dynamics revealed the presence of transverse waves propagating at relativistic speeds along the jet. Based on the correlated variations in the position angles of the QSC and the transverse wave, they proposed that the transverse waves are excited at the QSC, analogous to shaking the handle of a whip to generate transverse waves. In the ‘whip’ model, the jet is characterized by a strong helical magnetic field, where the transverse waves are identified as Alfvén waves excited at the QSC, while the moving components represent fast and slow magnetosonic compressions of the magnetic field and/or jet plasma.

\cite{arshakian20, arshakian24} conducted detailed investigations of QSC dynamics in the BL~Lac jet, demonstrating that the QSC trajectory exhibits reversal patterns, with apparent speeds frequently exceeding the speed of light. They proposed a model in which the observed reversing motions are projections of the QSC motion across the jet axis, akin to a seagull riding ocean waves. In the ‘seagull-on-wave’ model, the transverse waves originate upstream of the QSC location, with the QSC moving at subluminal speeds in the host galaxy frame. However, due to projection and relativistic effects, this motion appears to the observer as successive reversing patterns with superluminal speeds. Within this framework, the observed superluminal speeds of the QSC in blazars serve as evidence of the presence of relativistic transverse waves, with the scatter in QSC positions reflecting the amplitude of these waves.

Motivated by these findings in BL~Lac, we selected a different type of blazar for further study, the flat-spectrum radio quasar (FSRQ) 3C~279, known for its bright, compact radio core at 15~GHz \citep{lister19}. The latter is mainly due to the close alignment of the jet axis towards Earth, combined with the high Lorentz factor of the jet beam ($\theta \approx 2\degr$ and $\Gamma_{\rm beam} \approx 20$), which contributes to the observed characteristics of 3C~279 \citep[e.g.][]{pushkarev17}. Observations at 15~GHz reveal a QSC located at a median projected distance of 0.35~mas ($2.2$~pc) from the core \citep{lister19}. A bright component at about 0.4~mas is visible in the very high-angular-resolution images of the 3C~279 jet obtained with the space Very Long Baseline Interferometry (VLBI) mission \textit{RadioAstron} at 22~GHz \citep{fuentes23}, believed to correspond to the QSC observed at 15~GHz. At 43~GHz, the QSC appears at shorter separations of 0.25–0.3~mas \citep{wehrle01, jorstad04}, which could be due to the variability of this separation between the core and the QSC or to higher angular resolution that resolve the 15~GHz QSC into two features.

We conducted a detailed study of the on-sky distribution of QSC positions and QSC dynamics in the 3C~279 jet using over 25 years of 15~GHz VLBA monitoring data from the MOJAVE programme (Monitoring Of Jets in Active Galactic Nuclei with VLBA Experiments, Lister et al. 2009). Observational data and their uncertainties are presented in Section~\ref{sec:obs_data}. In Section~\ref{sec:on-sky_scatter}, we analyse the on-sky distribution of QSC positions, and in Section~\ref{sec:kinematics}, we examine the QSC kinematics. The characteristics of the QSC trajectory and transverse waves in the jet are estimated in Section~\ref{sec:characteristics}, followed by the discussion in Section~\ref{sec:discussions}. 

We assumed a flat cosmology with $\Omega_{m} = 0.27$, $\Omega_{\Lambda} = 0.73$, and $H_0 = 71$~km~ s$^{-1}$~Mpc$^{-1}$ \citep{komatsu09}. For the 3C~279 at redshift $z = 0.536$ \citep{marziani96}, the scale factor is 6.31~pc~mas$^{-1}$.

\section{Observational data: Quasi-stationary component}
\label{sec:obs_data}

For the purposes of our study, we used the publicly available VLBA observational data of quasar 3C~279\footnote{\url{https://www.cv.nrao.edu/MOJAVE/sourcepages/1253-055.shtml}} at 15~GHz obtained predominantly within the MOJAVE programme \citep{lister21} and its predecessor, the 2-cm VLBA Survey \citep{kellermann98,Zensus02}, with the rest from the NRAO data archive. The data spans from July 25 1995 through March 18 2022, encompassing a total of 167 high-quality observational epochs. The reconstructed brightness distribution maps from these observations typically exhibit a dynamic range of several thousand. 

\begin{figure}[hpt]
\centering
\includegraphics[width=9cm,angle=0] {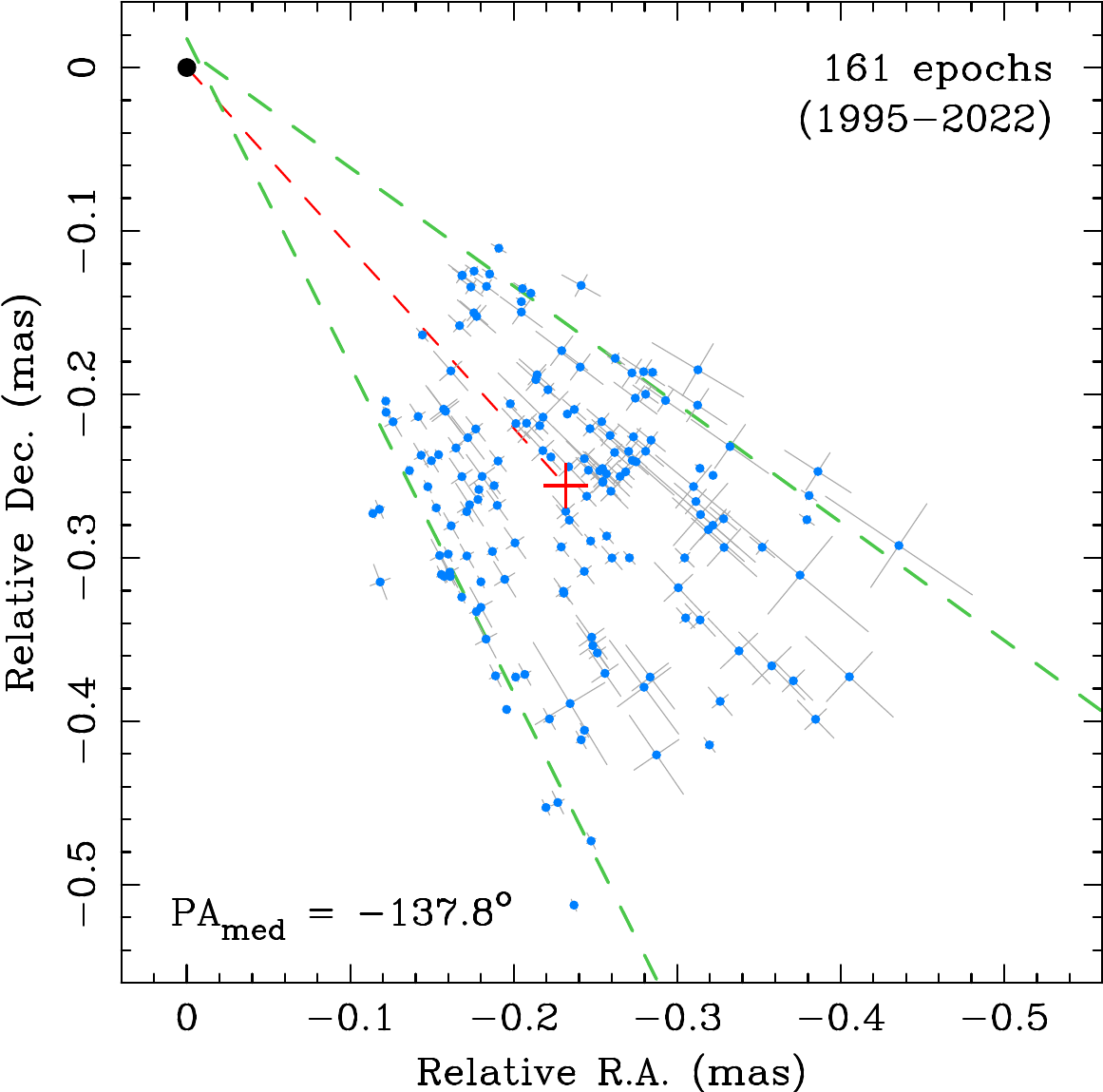}
\caption{Scatter of 161 positions of QSC on the sky plane. The sizes of the crosses correspond to QSC position errors in the directions toward of and across the core. The radio core is marked by a filled circle at position (0,0), and the median position of the QSC scattering is marked by a red plus sign. The dashed red line is the central axis of the jet, which connects the median positions of the QSC and the core. The dashed green lines represent the linear fit of the cone lines.} 
\label{fig:QSC_scatter_errors}
\end{figure}

Data reduction, including amplitude and phase calibration, was performed with the AIPS software package \citep{AIPS}, with subsequent imaging and structure model fitting in the Caltech Difmap package \citep{difmap} as was described in detail in \cite{lister09}. The modelfits up to June 29 2019 were taken from \cite{lister21}, while those at later epochs were obtained by us. The QSC is present at all but six epochs. Its position with respect to the VLBI core, defined as an apparent origin of the jet, is anisotropically scattered around the median location at $\sim 0.35$~mas from the core at a position angle PA$_{\rm {med}}\approx -138\degr$ (Fig.~\ref{fig:QSC_scatter_errors}). The average observing cadence is about 2 months, with a gap of 2.36 years between 2010.89 and 2013.25.

The position and flux density uncertainties of the fitted components were estimated following the approach presented in \cite{schinzel12}. However, this approach does not provide azimuthal dependence of the positional errors. To account for this, we inferred a median value of 0.57 for the ratio of the error across of the jet to that in the direction of the jet, following the procedure based on the $\chi^2$ statistics described in \cite{arshakian20} (see Sec.~3) and \cite{lampton76}.  We tested these theoretical position error estimates by randomly selecting the (u,v) coverage from 20 epochs and replacing the data in those epochs with known simulated structure based on real models from other randomly selected epochs. The simulated models used a combination of optically thin and thick spheres and varied in terms of the number and strength of components. We then conducted a blind modelfit to these simulated data and compared the fitted positions of the innermost component with the known simulated positions. The simulations showed good agreement with the positional errors obtained from the analytical approximations, but required an overall rescaling by 1.37 and additional rescaling to account for their asymmetry. 

\FloatBarrier 

The association of the core position with the conical shape of the QSC positions suggests that the latter is due to significant displacements of the core position along the jet axis. The QSC should move almost in a plane normal to the jet axis. In this case, displacements of the core position along the jet axis cause the intrinsic positions of the QSC to scatter downstream and upstream. If the viewing angle of the jet is larger than the intrinsic jet opening angle (i.e. we observe the jet outside its cone generatrix), we should observe an almost conical spread of positions. The size of the scatter in the direction of the jet reflects the intrinsic motion of the QSC and the motion of the core, while the size of the scatter across the jet through the median centre is due to the intrinsic motion of the QSC.

\begin{figure}[hpt]
\centering
\includegraphics[width=9.5 cm,angle=0] {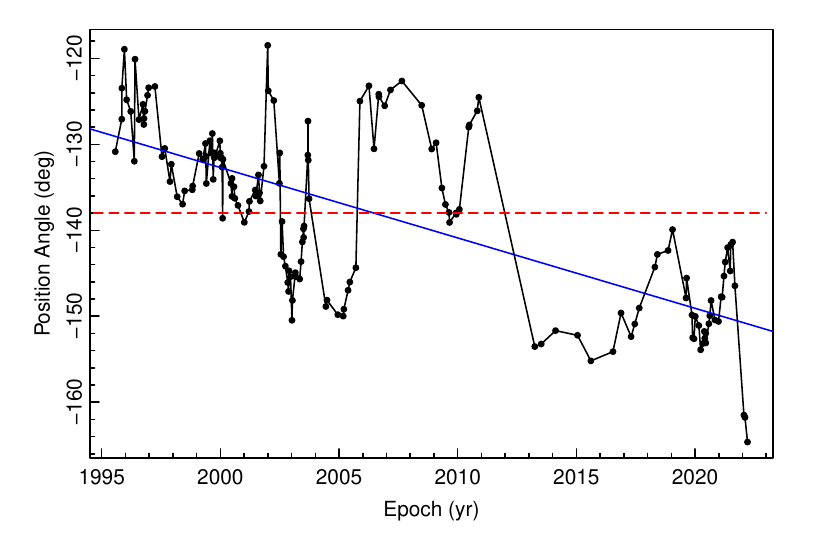}
\caption{Variation in QSC position angle with time. The dashed red line is the position angle of the jet central axis, ${\rm PA}_{\rm jet} = -138\degr$, and the blue line is a linear fit to the changes in QSC position angles with time. }
\label{fig:pa_qsc-epoch}
\end{figure}
\begin{figure*}[hpt]
\centering
\includegraphics[width=18cm,angle=0] {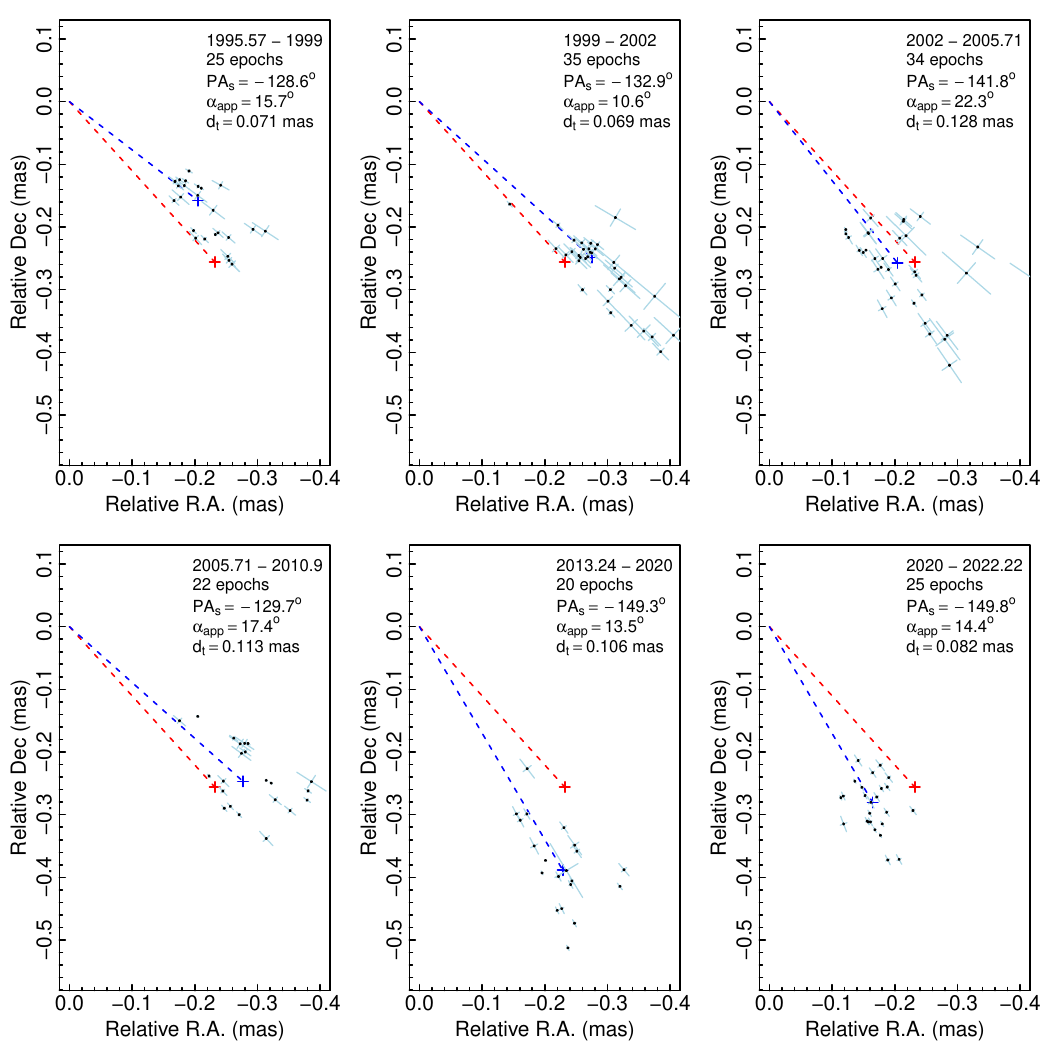}
\caption{Positions of the QSC within the six time intervals, 1995.57$-$1999, 
1999$-$2002, 2002$-$2005.71, 2005.71$-$2010.9, 2013.24$-$2020 and 2020$-$2022.22. The blue plus sign is the median position of the QSC positions within the given time interval. The dashed blue line is the axis of the jet connecting the core and the median position of QSC. The dashed red line is the central axis of the jet, connecting the median position of QSC over the entire observation period 1995.57$-$2022.22 and the core. The sizes of the crosses correspond to QSC position errors in the directions toward of and across the core.}
\label{fig:qsc_scatter-epoch}
\end{figure*}

\section{On-sky scatter of QSC positions}
\label{sec:on-sky_scatter}
The spread of QSC positions on the sky and their position uncertainties are shown in Fig.~\ref{fig:QSC_scatter_errors}. The median position error across and along the jet is $\sim 5$~$\mu$as and $\sim 12$~$\mu$as, respectively. The small positioning errors are attributed to high compactness and brightness of both the radio core and the QSC. The on-sky distribution of the QSC positions has a conical shape with cone lines  directed towards the core position at (0,0) (Fig.~\ref{fig:QSC_scatter_errors}). The size of the scatter along the jet axis is $\sim 0.3$~mas and across the jet at the position of the median centre is $\sim 0.15$~mas. To check if the cone lines intersect close to the core and measure the cone angle, we excluded three positions at $\rm{R.A.} > -0.12$~mas, which are positions of the last three observational epochs (see Fig.~\ref{fig:pa_qsc-epoch}) and stand out from the main scatter of positions. We used binning of positions in six non-overlapping bands transverse to the jet axis; for each band, we selected positions in the 85\% quintile on either side of the jet axis and performed a linear fit for the selected positions on either side. The intersection of the conical lines (the apex of the cone) is located very close to the core at a distance of 0.02~mas. According to our estimates, the cone angle is about $29\degr$. A change in the quantile from 75\% to 90\% results in a change in the angle of the cone that ranges from $26\degr$ to $32\degr$, respectively, and the distance between the core and apex of the cone within 0.05 mas. We assume that the conical shape of the position distribution is associated with the VLBI core and take the apparent full-cone angle to be $\phi_{\rm app}\approx 30\degr$. The latter corresponds approximately to the $\pm1.5$ standard deviation of the QSC position angles, $\sigma_{\rm PA_{\rm s}}$. (Further, all designations with subscripts $s$ or $c$ indicate the intrinsic characteristics of the QSC and the core, respectively). In the following, we will use the value $2 \times (1.5\sigma_{\rm PA_{\rm s}}) = 3\sigma_{\rm PA_{\rm s}}$ as a proxy for the apparent full cone angle.  

In the following, we examine the variation in QSC position angles with respect to the central jet axis (Fig.~\ref{fig:pa_qsc-epoch}). We note that the changes in PA$_{\rm s}$ are sensitive to QSC motions in directions transverse to the jet, and less sensitive to motions along the jet and hence to core displacements, which occur mainly along the jet axis. A linear fit of the position angles has a residual standard error of $7.1\degr$ and shows that PA$_{\rm s}$ decreases with time from about $-128\degr$ to $-150\degr$ over 27 years (blue line). It has an oscillatory behaviour and remains above median PA$_{\rm med} = -138\degr$ (dashed red line) until about $2001$, between 2001 and 2011 the amplitude of the PA oscillation increases and crosses the jet axis several times, and after 2013, variations in PA$_{\rm s}$ occur at position angles well below PA$_{\rm med}$. This is a clear indication that the jet axis changes direction with time. To quantify the change in the apparent full cone angle with time, we used the six specified time intervals $1995.57-1999$, $1999-2002$, $2002-2005.71$, $2005.71-2010.9$, $2013.24-2020$, and $2020-2022.22$. The choice of time intervals was due to the similarity of the changes in PA and the presence of a statistically significant number of PAs in each time interval ($\gtrsim 20$). For each time interval, we estimated the mean PA$_{\rm s}$ of the QSC, which represents the position angle of the jet PA$_{\rm jet}$, and the standard deviation $\sigma_{\rm PA_{\rm s}}$, which is then used to calculate the apparent full angle of the cone, $\phi_{\rm app} = 3\sigma_{\rm PA_{\rm s}}$ (see Fig.~\ref{fig:qsc_scatter-epoch}). 

The direction of the jet axis, as defined by PA$_{\rm s}$, exhibits a general trend of moving southwards. Initially, the jet axis moves clockwise until 2005.71, then reverses direction during the period 2005.71 to 2010.9, before resuming its clockwise motion towards the south. This behaviour may result from the superposition of two precessional motions: a long-term precession that drives the overall southward shift and a shorter-term precession (nutation) with a characteristic timescale of approximately five years (as observed between 2005.71 and 2010.9). Over a span of 27 years, the jet axis has changed its orientation on the sky by $21\degr$. If this variation is due to long-term jet precession, the corresponding precession angle and period must be at least $21\degr$ and $\gtrsim 40$ years, respectively. The position angles at the last three epochs suggest that the jet axis continues the long-term trend of its southward motion (Fig.~\ref{fig:pa_qsc-epoch}). Extended VLBA monitoring of 3C~279 at 15~GHz is required to test the jet precession hypothesis. The change in the PA of the inner jet towards a southerly direction was also revealed from the {\it RadioAstron} observations in March 2014 \citep{fuentes23} and February 2018 \citep{toscano25}.

The average apparent cone angle calculated over six time intervals is $\overline{\phi}_{\rm app} = 15.7\degr \pm 1.6\degr$. This estimate closely matches twice the residual standard error of the linear fit shown in Fig.~\ref{fig:pa_qsc-epoch}, with $15.7 \approx 2 \times 7.1 = 14.2$. Significant deviations of $\phi_{\rm app}$ from the mean value occur during the period 1999–2005.71. Our estimate of the mean apparent cone angle is consistent with other measurements. An apparent cone angle of $\phi_{\rm app} \approx 12\degr$ was measured by modelling the size of the core and the moving C4 component at 22 GHz \citep[][]{wehrle01}, using VLBA monitoring data from 1992 to 1998. This time interval partially overlaps with the period 1995.57–1999, during which we obtained $\phi_{\rm app} = 15\degr$. The apparent cone angles of $\phi_{\rm app} = 14\degr$ and $\phi_{\rm app} = 11.6\degr \pm 0.8\degr$ were also estimated using a single-epoch 15~GHz map \citep[][]{pushkarev09} and multi-epoch 15~GHz VLBA observations, respectively, with the latter based on deconvolution of the transverse jet width along the jet ridge line at distances $>0.5$ mas from the core \citep[][]{pushkarev17}. We argue that all these estimates are valid and that the differences between them are likely due to the use of data from different observational epochs and recognize that the apparent opening angle of the QSC cone tracks well with the apparent opening angle of the jet, as determined by the moving components and the transverse width of the jet.

The $\overline{\phi}_{\rm app} = 15.7\degr$, combined with the rotation of the jet axis by about $21\degr$, forms an apparent cone angle of $30\degr$ measured over the entire period of 27 years of observations. The deprojected intrinsic mean full-cone angle is then $\overline{\phi}_{\rm int}\approx 0.28\degr \pm 0.03\degr$ due to the foreshortening by a factor of approximately 60 in the case of the jet viewing angle $\theta = 1\degr$ (see the detailed discussion of the estimates of the jet viewing angle in Section~\ref{subsec:ssq}).

In the seagull-on-wave model, the QSC moves nearly across the jet axis due to the passage of a relativistic, swirled transverse waves \citep{arshakian24}, and the extent of the QSC scatter reflects the maximum amplitude of this wave. The transverse size of the opening cone at the median position, $d_{\rm t} = d_{\rm j}\, \phi_{\rm app}$, where $d_{\rm j}$ is the distance along the jet axis to the median position of QSC, can thus serve as an indicator of the maximum amplitude of the transverse wave in the jet, $A_{\rm w, max}$. For the six specified time intervals, we estimated $d_{\rm t} \approx A_{\rm w, max}$ values of 0.071, 0.069, 0.129, 0.114, 0.107, and 0.082~mas (Fig.~\ref{fig:pa_qsc-epoch}). The maximum amplitude of the transverse wave varies almost twice from 0.07~mas to 0.13~mas, with a mean maximum amplitude of $\overline{A}_{\rm w, max} = 0.10 \pm 0.01$~mas ($0.60 \pm 0.06$~pc).

\section{Kinematics of quasi-stationary component}
\label{sec:kinematics}
\cite{arshakian24} proposed a seagull-on-wave model of the QSC and showed for the BL~Lac jet that the measured superluminal velocities of the QSC indicate the presence of relativistic transverse waves travelling downstream through the QSC. In this section, we examine the apparent displacements $r$ passed by the QSC during the time, $\Delta t$, between consecutive observing epochs and the apparent speeds $\beta_{\rm r} = \mu D_{\rm A}(z)(1+z)/c$ expressed in units of the speed of light ($c$), where $\mu = r/\Delta t$ is the proper motion of the QSC and $D_{A}(z)$ is the angular diameter distance at redshift $z$. In fact, the apparent displacement vector of the QSC ($\vec{r}$) is a combination of the core displacement vector ($\vec{c}$) and the intrinsic displacement vector of the QSC ($\vec{s}$) \citep{arshakian20}. A careful analysis of the apparent displacements and observation time intervals is necessary to disentangle the statistics of intrinsic displacements of the QSC and the core.   

\begin{figure}[hpt]
\centering
\includegraphics[width=9.5 cm,angle=0] {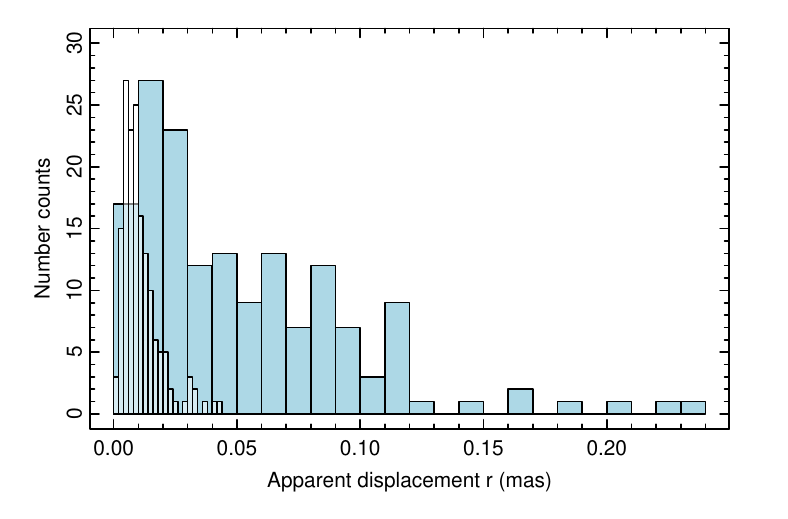}
\caption{Distributions of 160 apparent QSC displacements (light blue area) and their uncertainties (transparent).}
\label{fig:disp_dist}
\end{figure}

\subsection{Apparent displacements} In VLBA maps at 15~GHz, a compact and bright radio core is used as a reference point to measure the distances to the jet features. Due to the use of the phase self-calibration procedure for reconstructing VLBI images, the absolute position information is lost. Therefore, if the core itself is moving and the QSC has its own intrinsic motion, the observer measures the apparent motion of the QSC, which is a combination of the intrinsic displacements of the core and the QSC. At centimetre wavelengths, the core corresponds either to the section of the jet where the optical depth is close to unity or to a stationary feature(s) (or a combination of both) \citep{marscher08ASPC}. For a synchrotron self-absorbed core, displacements are expected to occur along the jet axis, driven by synchrotron opacity effects, as well as variations in particle density and magnetic field strength. Stationary features located near the core may appear shifted from the jet axis owing to the small viewing angle. When blended with the self-absorbed core, this can result in apparent core displacements with a transverse component relative to the jet direction.

To study the dynamics of the QSC, we introduced the apparent displacement vector, $\vec{r}$, where the direction of the vector and its magnitude, $r=\left|{\vec{r}}\right|$, were determined by the position of the QSC between successive epochs of observations, $t$ and $t+\Delta t$, where $\Delta t$ is the time interval between successive observations. The distribution of apparent displacements has a median value of $37\pm3.5~\mu$as, reaches a maximum at low values, and gradually decreases towards high values (Fig.~\ref{fig:disp_dist}). The errors of the apparent displacements, $\delta_{\rm r}$, were estimated using the error propagation method using the uncertainties of two consecutive positions \citep[see][]{arshakian20}. The median error is $\widetilde{\delta_{\rm r}}= 3.6~\mu$as. We excluded the six outliers beyond 0.15~mas from further analysis. We note that an outlier with $r=0.227$~mas is measured at the longest time interval (observational gap) of 2.36 years.

\begin{figure}[hpt]
\centering
\includegraphics[width=9.5 cm,angle=0]{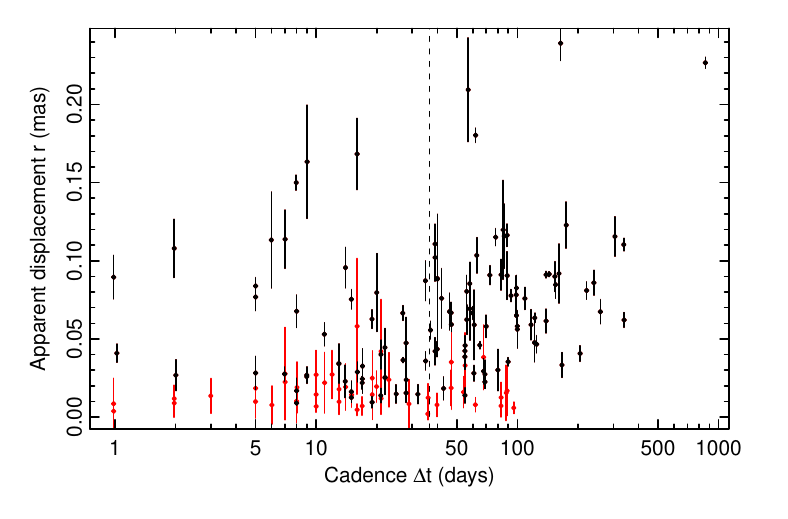}
\caption{Apparent QSC displacements as a function of the observation time interval. The $1\sigma$ uncertainties of the apparent displacements are presented. Apparent displacements with high relative errors $\varepsilon_{\rm r} > 0.5$ are indicated in red. The dashed vertical line indicates the median observational interval of 37 days. }
\label{fig:disp-dt}
\end{figure}

The apparent displacements are measured over a wide range of observation intervals, from one day to about one year (Fig.~\ref{fig:disp-dt}), with a median observational interval of 37 days. The apparent displacements with high relative errors, $\varepsilon_{\rm r} = \delta_{\rm r} / r > 0.5$, which are considered unreliable, are distributed in the lower part of the relation in the range of observational intervals from 5 to 100~days (blue colour). Those with $\varepsilon_{\rm r} < 0.5$ have a natural tendency to decrease as the observation interval shortens. The typical positional uncertainty in Fig. \ref{fig:disp_dist} is 0.01 mas, meaning that we should expect displacements between two epochs to be typically of order 0.014 mas with values tailing up to three times this due to uncertainty alone.  The vast majority of short cadence displacements are there likely due to simply to uncertainty, although a few of the larger displacements may be real.  We caution however that large displacements over very short time intervals are suspect due both to uncertainty in the displacements themselves and in the time intervals which are accurate at best to 0.5 days. 
\FloatBarrier 
\begin{figure}[hpt]
\centering
\includegraphics[width=9.5cm,angle=0] {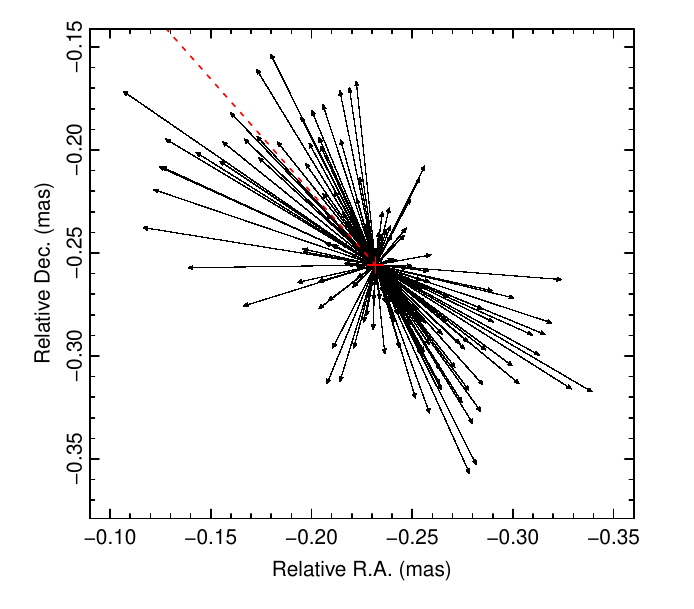}
\caption{Selected 154 apparent displacement vectors translated to the median QSC position. The latter is marked by a plus red sign. The dashed red line is the jet central axis which connects the median position of QSC and core.}
\label{fig:shifted_disp}
\end{figure}

To study the orientation of apparent displacement vectors, we translated the displacement vectors so that their initial points coincide with the median centre of the QSC positions (Fig.~\ref{fig:shifted_disp}). Vectors of apparent QSC displacements show preferential orientation and larger magnitudes along the jet axis, very similar to the angular distribution of apparent displacements of BL~Lac \citep{arshakian20}. We suggest that the displacement asymmetry is predominantly caused by core wobbling, which occurs along the jet axis. This effect is mainly driven by variations in synchrotron opacity as new disturbances pass through the VLBI core and clearly dominates over any possible transverse displacements of the core.

\begin{figure}[hpt]
\centering
\includegraphics[width=9cm,angle=0] {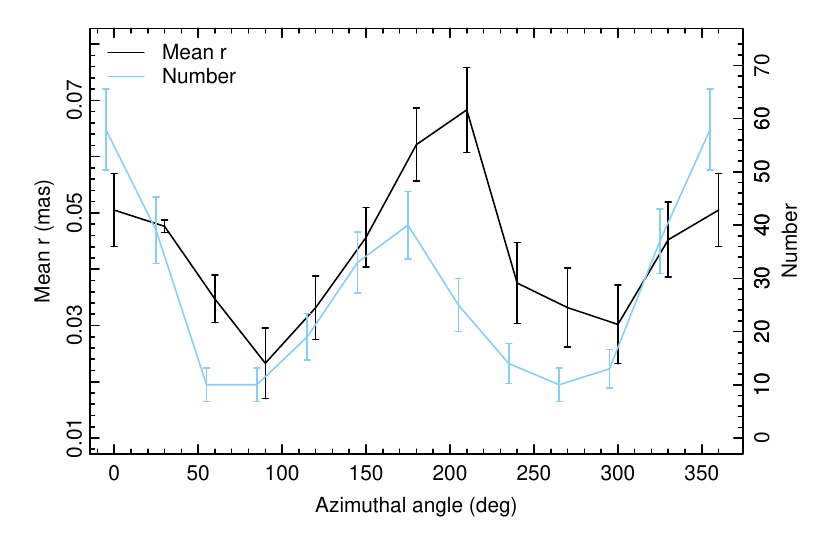}
\caption{Azimuthal distribution of the mean length and the number of displacement vectors (black and blue lines, respectively). }
\label{fig:azim_dist}
\end{figure}

To quantify displacement asymmetry, we considered azimuthal distributions of the lengths and number of displacement vectors. The azimuthal angle is the angle between the displacement vector and the downstream direction of the jet axis, which increases anticlockwise from $0\degr$ to $360\degr$. The distribution of the mean displacement lengths was calculated within an angular beam of $60\degr$ in increments $30\degr$, and the mean length errors were calculated using Eq.~(1) in the \cite{arshakian20}. The distribution of the mean lengths of displacements is significantly asymmetric, the lengths are larger along the jet ($\sim 0.6$~mas), and on average these are about three times smaller ($\sim 0.2$~mas) in direction across the jet (Fig.~\ref{fig:azim_dist}). The change in the number of vectors with azimuthal angle is similar to the behaviour of the mean length (blue line). In general, the asymmetry and anisotropy of the apparent displacement vectors can be interpreted as the effect of asymmetric core displacements. The variation in the number of vectors shows a significant difference in the jet direction and across to it. Moreover, the number of downstream vectors is larger by a factor of 1.5 than those upstream the jet. This is likely related to the apparent motion of the core caused by disturbances and/or inhomogeneities propagating through it. The apparent oscillatory motion of the core relative to its undisturbed position is asymmetric, typically characterized by a rapid rise in flux density (corresponding to upstream displacement vectors of the QSC as the core shifts downstream), followed by a slower decay phase (associated with downstream displacement vectors of the QSC as the core gradually returns upstream to its undisturbed position).

\paragraph{Intrinsic characteristics.} Once we establish that the core displacements occur mainly along the jet axis, one can use the statistical method developed in \cite{arshakian20} to unravel the apparent displacements of QSC. It allows for the evaluation of intrinsic characteristics of the core and QSC such as the mean, standard deviation, and rms (root mean square) of intrinsic displacements using projections of apparent displacements on and across the jet axis ($r_{\rm j}$ and $r_{\rm n}$, respectively; see their Eqs.~(9,10),~(13,14), and (17,18)). The apparent displacements that occur over small observation time intervals should display more realistic statistics of the intrinsic displacements of QSC. To examine the latter, we used the 109 apparent displacements with $\varepsilon_{\rm r} < 0.5$ and considered subsamples of apparent displacements constrained to cadence ranges of 0–14, 14–33, 33–60, 60–100, and 100–365 days, ensuring that the number of apparent displacements remained close to 20 (see Table~\ref{table:2}). For the interval of 14–33 days, the intrinsic apparent displacement reaches its minimum ($\overline{s} \approx 0.02$~mas) and the corresponding ratio $\rm rms_c / \rm rms_s = 1.62$.

\begin{table*}[htbp]
\caption{Estimates of intrinsic characteristics of the QSC and radio core for specified cadence ranges $\Delta T$, with relative errors $\varepsilon_{\rm r} < 0.5$.}
\label{table:2}
\centering
\begin{tabular}{ccccccccc}
\hline \hline 
\multicolumn{1}{c}{$\Delta T$} & \multicolumn{1}{c}{$N$} & \multicolumn{1}{c}{$\overline{s}$} & \multicolumn{1}{c}{$\sigma_{\rm s}$} & \multicolumn{1}{c}{${\rm rms}_{ \rm s} \pm \delta_{\rm rms_{ \rm s}}$} & \multicolumn{1}{c}{$\overline{c}$} & \multicolumn{1}{c}{$\sigma_{ \rm c} = {\rm rms}_{ \rm c} \pm \delta_{\rm rms_{ \rm c}}$} & \multicolumn{1}{c}{$\overline{\Delta t} \pm \sigma_{\rm \overline{\Delta t}}$} & \multicolumn{1}{c}{$\overline{\beta_{\rm s}} \pm \sigma_{\rm \overline{\beta_{\rm s}}}$} \\ 
\multicolumn{1}{c}{(days)} & \multicolumn{1}{c}{} & \multicolumn{1}{c}{(mas)} & \multicolumn{1}{c}{(mas)} & \multicolumn{1}{c}{} & \multicolumn{1}{c}{(mas)} & \multicolumn{1}{c}{(mas)} & \multicolumn{1}{c}{(days)} & \multicolumn{1}{c}{(c)} \\ 
\multicolumn{1}{c}{(1)} & \multicolumn{1}{c}{(2)} & \multicolumn{1}{c}{(3)} & \multicolumn{1}{c}{(4)} & \multicolumn{1}{c}{(5)} & \multicolumn{1}{c}{(6)} & \multicolumn{1}{c}{(7)} & \multicolumn{1}{c}{(8)} & \multicolumn{1}{c}{(9)} \\ 
\hline
0 - 14 & 21 & 0.034 & 0.010 & $0.035 \pm 0.007$ & $-0.004$ & $0.063 \pm 0.002$ & $7.41 \pm 4.05$ & $52.42 \pm 7.16$ \\ 
14 - 33 & 22 & 0.019 & 0.010 & $0.021 \pm 0.008$ & $-0.004$ & $0.034 \pm 0.004$ & $21.49 \pm 5.10$ & $10.00 \pm 1.26$ \\ 
33 - 60 & 23 & 0.026 & 0.016 & $0.030 \pm 0.014$ & $0.017$ & $0.060 \pm 0.010$ & $46.79 \pm 8.49$ & $6.38 \pm 0.86$ \\ 
60 - 100 & 22 & 0.038 & 0.019 & $0.043 \pm 0.008$ & $-0.005$ & $0.066 \pm 0.004$ & $80.38 \pm 13.38$ & $5.48 \pm 0.62$ \\ 
100 - 365 & 21 & 0.052 & 0.015 & $0.054 \pm 0.007$ & $-0.013$ & $0.058 \pm 0.006$ & $182.61 \pm 74.12$ & $3.29 \pm 0.36$ \\ 
\hline
\end{tabular}
\tablefoot{Columns are as follows: (1) $\Delta T$ is the range of observation time intervals (or cadences), (2) number of apparent displacements, (3) $\overline{s}$ is the mean intrinsic apparent displacement of QSC, (4) $\sigma_ {\rm s}$ is the standard deviation of the intrinsic displacements, (5) ${\rm rms}_{\rm s}$ of the intrinsic displacements of QSC and its standard error, (6) $\overline{c}$ is the mean displacement of the core, (7) ${\rm rms}_{\rm c}$ of the displacements of the core and their standard errors, (8) mean and standard deviation of the cadences, (9) mean intrinsic apparent speed of QSC $\overline{\beta_{\rm s}}=\overline{s} / \overline{\Delta t}$ and its standard deviation.
}
\end{table*}

If the motion of the QSC is quasi-linear, we would expect the mean intrinsic apparent displacement, $\overline{s}$, to decrease with shorter time intervals. Indeed, $\overline{s}$ decreases to a minimum value of $\approx 0.02$~mas as the time interval decreases to $\Delta T = 14-33$~days, and then increases to 0.034~mas for even shorter intervals (Table~\ref{table:2}). The latter is most likely due to uncertainties in the displacements themselves and the limited accuracy of time interval measurements of order half a day (Fig.~\ref{fig:disp-dt}). Another explanation is that the QSC motion becomes non-linear on characteristic timescales of $\lesssim 10$~days. These are a subject for a separate study; here, we adopt $\Delta T = 14-33$~days, for which the intrinsic apparent displacement reaches its minimum value of $\overline{s} = 0.019$~mas. The rms displacement of the core is 1.62 times larger than that of the QSC, indicating that the anisotropic motion of the core dominates over the intrinsic motion of the QSC. At longer time intervals, up to $\lesssim 100$~days, this ratio fluctuates within the range 1.6–2 (see Table~\ref{table:2}).

We attempted to analyse the variation in the rms of the core and QSC with epoch. However, after filtering displacements for small time intervals, the statistical samples per epoch bin became insufficient to draw a definitive conclusion.

\subsection{Superluminal speeds of QSC}
\label{subsec:ssq}
Regarding the intrinsic apparent speeds of the QSC, we can determine the mean intrinsic speed as $\overline{\beta}_{\rm s} = \overline{s} / \overline{\Delta t}$, where $\overline{s}$ represents the mean intrinsic displacement of the QSC, and $\overline{\Delta t}$ denotes the mean observation time interval. The smaller the observation time interval, the more accurate the estimate of the intrinsic apparent displacement of the QSC. We used the values of $\overline{s} = 0.019$~mas and $\overline{\Delta t} = 21.5$~days, obtained for the range of $\Delta T = 14-33$~days, to estimate the mean intrinsic apparent speed of the QSC, $\overline{\beta}_{\rm s} \approx 10 \pm 1.3$ (see Table~\ref{table:2}). The superluminal speed of the QSC can be explained by the seagull-on-wave model, in which a relativistic transverse wave propagates through the stationary QSC, displacing it in a direction transverse to the central axis of the jet \citep{arshakian24}. If the relativistic transverse wave moves in a direction nearly aligned with the line of sight, the subluminal speed of the QSC in the host galaxy's rest frame may appear superluminal in the observer's rest frame. The fact that all measured mean intrinsic apparent speeds of the QSC are superluminal (see Table~\ref{table:2}) provides strong evidence of the presence of a relativistic transverse wave in the 3C~279 jet.

We now attempt to estimate the characteristics of the relativistic transverse wave. \cite{arshakian24} derived a formula (their eq.~3) that describes the relationship between the characteristics of the QSC and the transverse wave in the reference frames of the host galaxy and the observer. This formula applies to the case where the wave propagates in the sagittal plane, the plane containing both the line of sight and the central axis of the jet:
\begin{equation}
    \beta_{\rm s} = \frac{\beta^{g}_{\rm wave}\tan\varphi \cos\theta}{(1-\beta^{g}_{\rm wave}\cos\theta)},
    \label{eq:beta_s_sagittal}
\end{equation}
\begin{equation}
    \tan{\varphi} = \frac{\beta^g_{\rm s,tr}}{\beta^g_{\rm wave}} ,  
    \label{eq:tg_varphi}
\end{equation}
where $\beta_{\rm s}$ is the intrinsic apparent speed of QSC as measured by observer, $\theta$ is the jet viewing angle, $\beta^g_{\rm s,tr}$ and $\beta^{g}_{\rm wave}$ are the intrinsic speed of the QSC and transverse wave propagation speed in the rest frame of the host galaxy, and  $\varphi$ is the wave inclination angle at the QSC position with respect to the jet propagation direction. They argue that these equations are also valid for waves propagating out of the sagittal plane if the jet viewing angle is less than approximately $10\degr$. The jet viewing angle of 3C~279 is estimated to be $\lesssim 4\degr$. \citet{piner03} use 22 and 43~GHz VLBI data and report viewing angles ranging from $2\degr$ to $4\degr$, depending on the distance of the component from the jet core. \citet{rani18}  estimate a jet viewing angle of $\le 2.6\degr$ and Lorentz factor of the flow $\ge 22.4$, based on parsec-scale collimated jet observations and variability analysis at 43~GHz, while \citet{pushkarev17} derive a viewing angle of $2.4\degr$ using apparent jet speeds \citep{lister16} and Doppler factors derived from 15~GHz flux density variability \citep{hovatta09}. Based on VLBA polarimetric observations and Doppler boosting, \citet{jorstad04} use 43~GHz VLBI data to estimate a very small viewing angle of $\approx 0.5\degr$. An analysis of 15~GHz data for the moving component C4, based on its observed kinematics and brightness evolution, suggests a required viewing angle of less than $1\degr$ and a Lorentz factor of $\Gamma > 15$ \citep{homan03}. Moreover, later robust components clearly reach apparent speeds of nearly $30c$, implying a Lorentz factor of at least $\Gamma \gtrsim 30$ \citep{homan15}. 
The range of viewing angles could be a result of varying the jet direction as evident from changes of PA$_{\rm s}$ by about $20\degr$ over 27 years, and that it may undergo nutation on shorter timescales (see Section 3), while the high Lorentz factors could be due to parsec-scale acceleration and/or bending the jet flow.
The bulk Lorentz factor of the jet in 3C~279 is typically estimated to lie in the range $\Gamma_{\rm beam} \sim 10$–$20$ \citep{lister13}, with most studies converging around $\Gamma_{\rm beam} \sim 12$–$15$ \citep{jorstad04, homan03, piner03}. In this study, we adopt a viewing angle of $\theta = 1\degr$, as smaller angles would imply a jet opening angle wider than observed \citep{pushkarev17}. We also assume a bulk Lorentz factor of $\Gamma_{\rm beam} = 30$, which corresponds to a maximum apparent speed of $25c$ for moving components at $\theta = 1\degr$. Under these assumptions, the wave speed is constant and the inclination angle of the transverse wave remains unchanged over the observation time interval.

We adopt $\overline{\beta}_{\rm s} \approx 10 \pm 1.3$ and $\sigma_{\beta_{\rm s}} \approx 8$. To estimate the apparent speeds of transverse waves in the jet of 3C~279, a detailed analysis of the jet ridge motion is required, following the approach outlined in \cite{cohen14,cohen15}. Here, we estimate the approximate range of wave speeds in 3C~279 by utilizing the relationship between the bulk speed and transverse wave speed observed in the jet of BL~Lac. In highly magnetized jets, where the magnetic field energy density exceeds the plasma energy density, the Alfvén speed can approach the speed of light, allowing transverse Alfvén waves to propagate faster than the jet's bulk flow \citep[e.g.][]{cohen15}. Assuming the presence of strong magnetic fields of the jet at the QSC distance, we assume that the Lorentz factor of the wave $\Gamma_{\rm w} \gtrsim \Gamma_{\rm beam} \approx 30$. The upper limit of the $\Gamma_{\rm w}$ of 3C~279 can be adopted from measurements of transverse patterns speeds on the BL~Lac jet. \cite{cohen15} provides a refined estimate of the Lorentz factor for the beam as $\Gamma_{\rm beam} \approx 4.5$ of BL~Lac. Additionally, superluminal speeds for Alfvén waves range between apparent velocities $\beta_{\rm{app}} \approx 3.9$ and $\beta_{\rm{app}} \approx 13.5$, with corresponding galaxy-frame speeds approximating $\beta^{g}_{\rm wave} \approx 0.98 - 0.998$ (or $\Gamma_{\rm w} \approx 5 - 15$), indicating that the pattern speed in BL~Lac is approximately two or three times faster than the beam speed, $\Gamma_{\rm w} \approx \Gamma_{\rm beam} - 3\Gamma_{\rm beam}$. Adopting the later approximation for the 3C~279 we obtain a range of the speeds of transverse patterns that is $\Gamma_{\rm w} \approx 30 - 90$, where the mean speed is $\overline{\Gamma}_{\rm w} \approx 60$ and standard deviation of the speeds $\sigma_{\Gamma_{\rm w}} \approx 10$.

\begin{figure}[hpt]
\centering
\includegraphics[width=9. cm,angle=0] {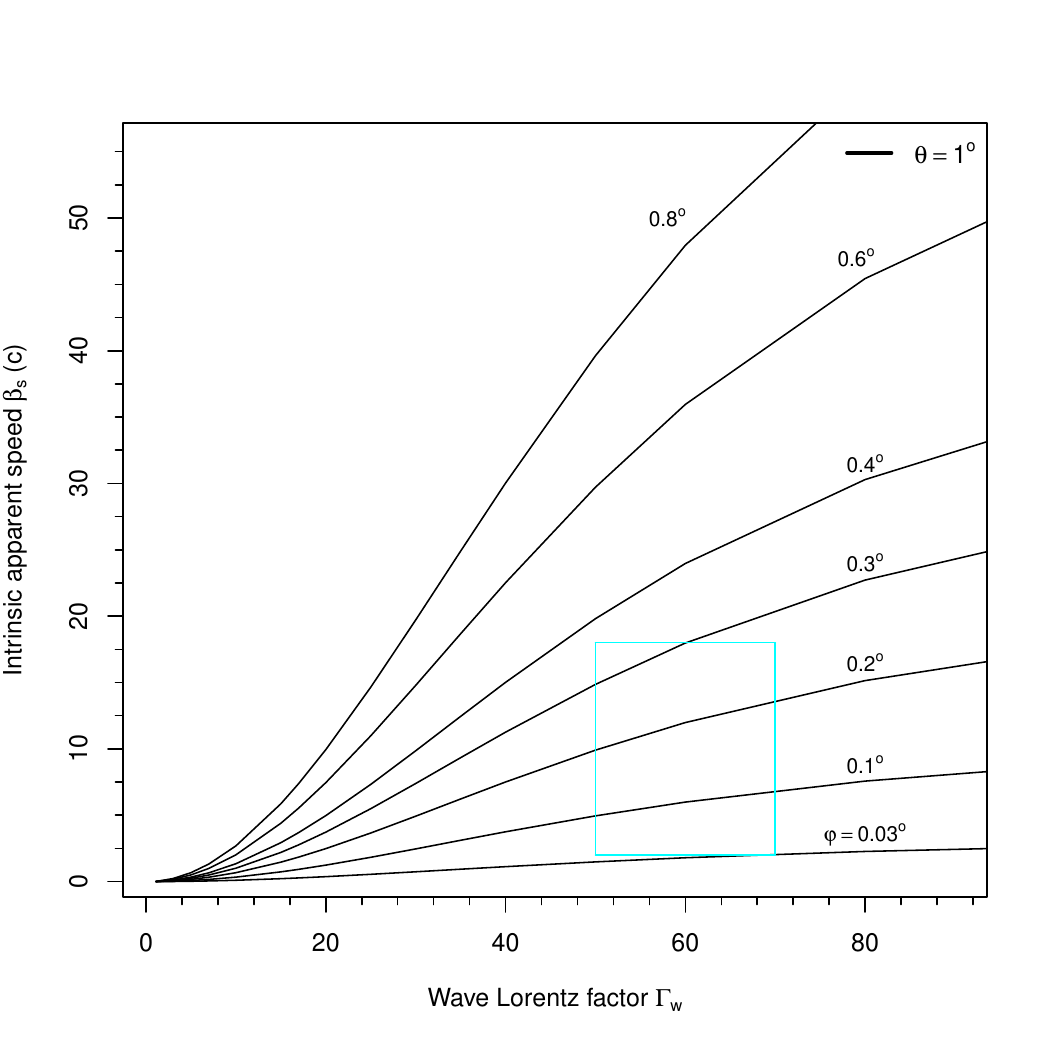}
\caption{Apparent transverse velocity of the QSC as a function of the Lorentz factor of the wave in case of the jet viewing angle $\theta = 1\degr$. Different lines correspond to the wave inclination angles $\varphi = 0.03\degr, 0.1\degr, 0.2\degr, 0.3\degr, 0.4\degr, 0.6\degr, 0.8\degr$. The area bounded by the cyan rectangle represents the $\pm 1\sigma$ deviations of $\beta_{\rm s}$ and $\Gamma_{\rm w}$ from the mean values $\overline{\beta}_{\rm s} = 10$ and $\overline{\Gamma}_{\rm w} = 60$, respectively.}
\label{fig:app_speed-gamma_w}
\end{figure}

In Fig.~\ref{fig:app_speed-gamma_w} we plot the relation between $\beta_{\rm s}$ and $\Gamma_{\rm w}$ (Eq.~\ref{eq:beta_s_sagittal}) for a fixed $\theta = 1\degr$ and different wave inclination angles ranging from $\varphi \approx 0.03\degr - 0.8\degr$ (black lines). The width of the rectangle (cyan colour) represents the range $\sigma_{\Gamma_{\rm w}} = \pm 10$ from $\overline{\Gamma}_{\rm w} \approx 60$ and the length is the range $\sigma_{\beta_{\rm s}} = \pm 8$ with respect to the mean intrinsic apparent speed of QSC, $\overline{\beta}_{\rm s} = 10$. From these statistics we estimated the mean inclination angle of the wave $\overline{\varphi} \approx 0.2\degr$ (Eq.~\ref{eq:beta_s_sagittal}) and defined the $\varphi$ limits confined by the rectangle, $\approx 0.03\degr$ and $\approx 0.4\degr$ (see Fig.~\ref{fig:app_speed-gamma_w}), which represent $\pm 1\sigma$ deviations from the mean inclination angle. This gives $\sigma_{\rm \varphi} \approx 0.2\degr$ and the upper limit of $\varphi \lesssim 0.8\degr$ estimated at the level $3\sigma$. It is important to note that these are rough, model-dependent estimates that may vary with changes in the jet viewing angle, the intrinsic speed of the QSC, and the properties of the transverse wave. For example, \cite{homan21} report a very high Doppler boosting factor of 140 and a Lorentz factor of 70 for the jet of 3C~279. If accurate, this would imply an average wave Lorentz factor of $\overline{\Gamma}_{\rm w} \approx 120$, which, in combination with $\overline{\beta}_{\rm s} \approx 10$, places an upper limit on the wave inclination angle, $\varphi \lesssim 0.4\degr$. 

\section{Characteristics of relativistic transverse waves}
\label{sec:characteristics}
In this section, we analyse the QSC trajectory to evaluate the transverse wave frequency and amplitude. The observed trajectory of the QSC is composed of apparent displacements, $r$, between successive positions of the QSC. The apparent displacement is a combination of the intrinsic apparent displacement of the QSC ($s$) and the displacement of the core ($c$). To recover the intrinsic trajectory of the QSC, it is necessary to smooth out the anisotropic displacements of the core, which occur along the jet axis. To achieve this, we apply a moving average procedure to QSC positions, thereby mitigating the effect of core displacement. In this method, a time window centred on each epoch rolls over the epochs and calculates the average position and its associated error within the smoothing time window (\( \Delta t_{\rm opt} \)). Edge treatment: the method excludes epochs from the moving average calculation when the full time window cannot be applied.

\paragraph{The length of the smoothing window.} The time interval for the sliding window should be chosen optimally to average out the effects of core displacement while preserving, as much as possible, the intrinsic trajectories of QSC. The length of time window should be large enough to smooth out wiggling of the core due to physical reasoning such as particle density and magnetic field changes that may result from the passage of moving plasma perturbations through the core \citep{arshakian24}. 

\begin{figure}[hpt]
\centering
\includegraphics[width=8.8cm,angle=0] {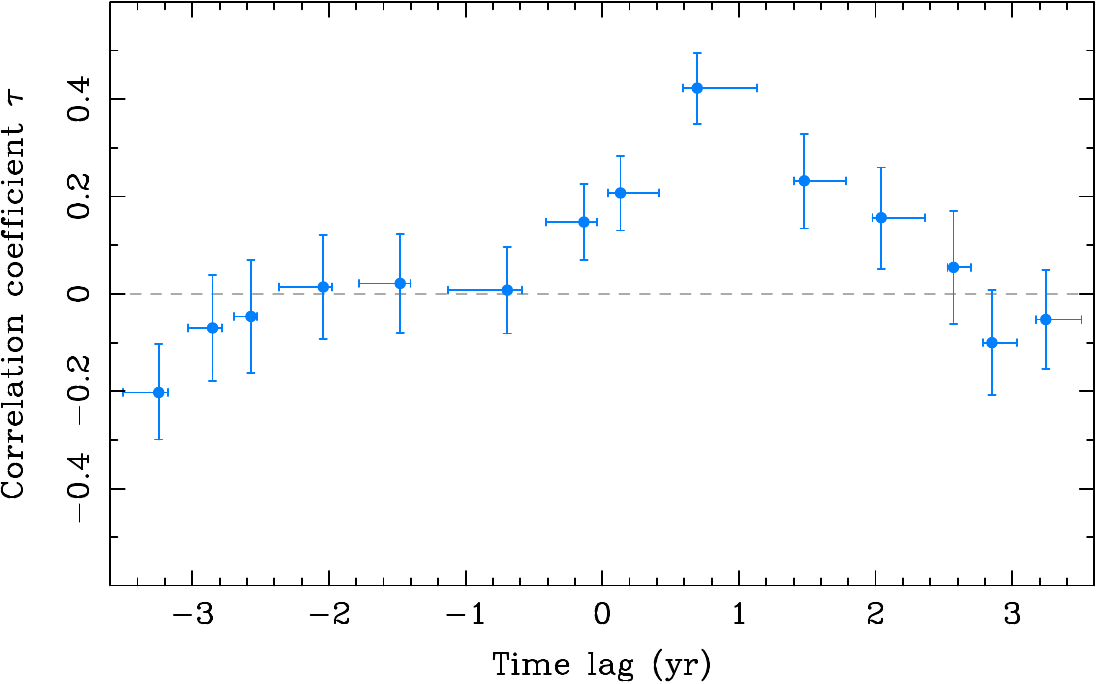}
\caption{Cross-correlation coefficient as a function of time lag for the flux density curves of the QSC and the radio core.}
\label{fig:zdcf}
\end{figure}

\cite{plavin19} suggested that moving components, such as shocks or plasma blobs, can influence the observed core position and brightness. These moving components can lead to temporary increases in core brightness and shift the apparent core position upstream when approaching the core and downstream, in direction of its motion, when moving away from the core. In the same way, moving components can cause apparent displacements of the QSC position when they pass through the QSC region. This will lead to correlated displacements of the core and QSC with a time delay equal to the time that the perturbation spent travelling the distance between the core and QSC. To estimate the time lag we applied the z-transformed discrete cross-correlation function \citep[zDCF;][]{alexander97} to variation in the flux density of the core and QSC with epoch. We found that the maximum correlation $\tau = 0.42$ reaches at a time lag of about 0.7~yr, where the variation in the core flux density is leading (Fig.~\ref{fig:zdcf}). We assume that the length of the sliding time window should be $\gtrsim 0.7$~years to properly smooth the correlated displacements of the core and QSC. 

Another issue is how often the disturbance passes through the core, i.e. the ejection rate of moving components. The quasar 3C~279 has been extensively studied to determine the ejection rate of moving components. \cite{jorstad01} monitored 3C~279 with VLBA at 22~GHz and 43~GHz over several years and identified multiple superluminal components. They estimated an average ejection rate of approximately one new component per year and reported variability in the ejection rate, with periods of increased activity where multiple components were ejected within a single year, followed by quieter periods with fewer or no ejections. We assume that the length of the smoothing window should be equal or longer than the timescales of the core disturbances due to moving components $\Delta t_{\rm opt} \gtrsim 1$~yr.  

\cite{arshakian24} noted that another possible bias in the fitted position of the components, potentially leading to apparent regular variations in the position of QSC in the BL~Lac jet, may arise from the model fitting process. This is not the case with 3C~279 since the core is very bright and compact with a relatively high median cadence of observations of about 40 days.

\begin{figure}[hpt]
\centering
\includegraphics[width=9.5cm,angle=0] {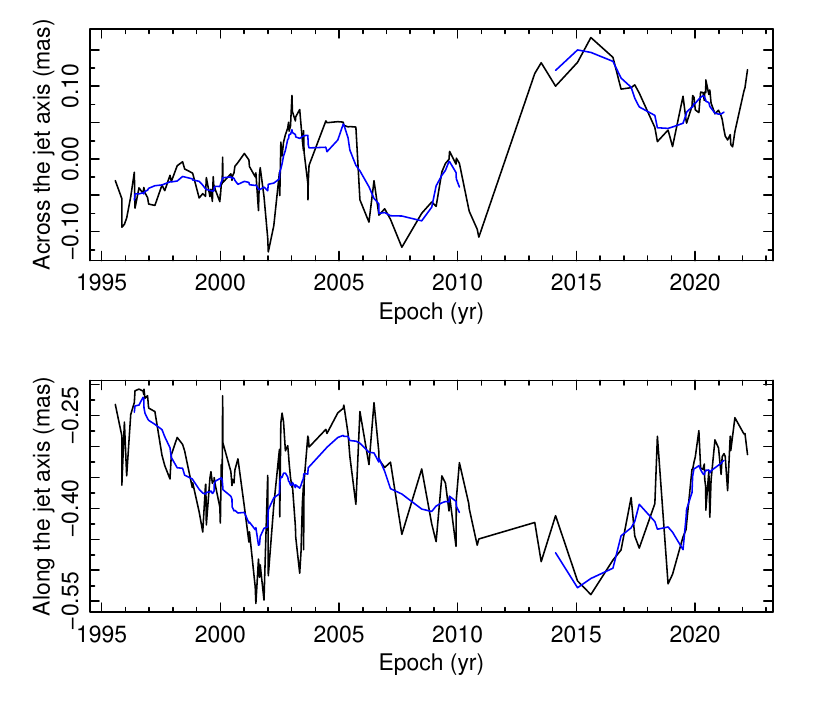}
\caption{Variations in the QSC position across and along the jet axis are shown by the black lines. The blue lines represent the smoothed curve obtained using the moving average method with a time window of 1.5 years.}
\label{fig:pos-epoch}
\end{figure}

Finally, we conducted a decision analysis of the QSC position displacements along and across the jet axis (black lines in Fig.~\ref{fig:pos-epoch}) to determine the optimal width of the smoothing window. The variations along the jet axis are more pronounced due to the core's oscillatory displacements in the direction of the jet. To effectively smooth these variations, the sliding window length must be sufficiently large to capture long-term variations on the order of one year (bottom panel). We adopt $\Delta t_{\rm opt} = 1.5$~yr, which effectively smooths variations along and across the jet axis on timescales of one year or longer, as illustrated by the blue lines in Fig.~\ref{fig:pos-epoch}.

\begin{figure}[hpt]
\centering
\includegraphics[width=9.5cm,angle=0] {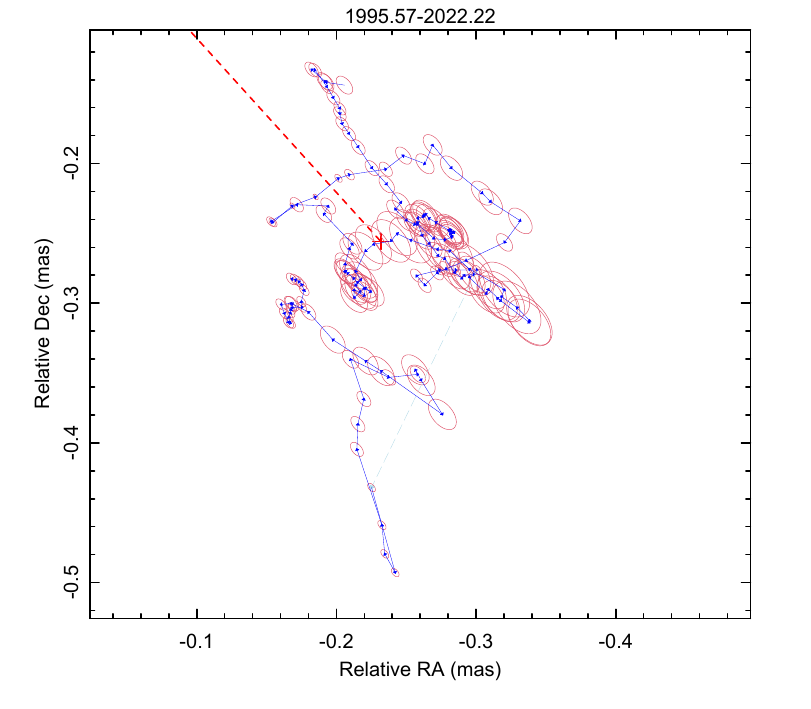}
\caption{Trajectory of the QSC from 1995.57 to 2022.22, smoothed using a time window of 1.5 years (blue line). The dashed light blue line connects the endpoints of the observation gap. Red ellipses indicate the average asymmetric positioning uncertainties. The median scatter position is marked with a red plus sign, while the dashed red line represents the central jet axis, connecting the median QSC position to the radio core.}
\label{fig:trajectory_smooth}
\end{figure}

\begin{figure*}[hpt]
\centering
\includegraphics[width=17cm,angle=0] {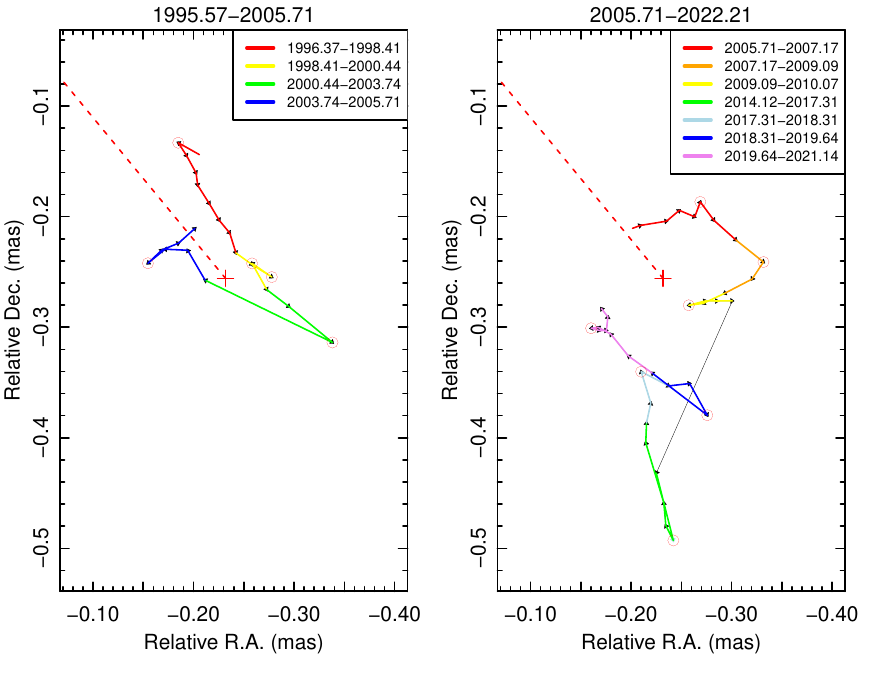}
\caption{Refined trajectory of the QSC represented for two time periods: 1995.57$-$2005.71 and 2005.71$-$2022.22. The arrows denote the direction of movement, and the red circles mark reversal point. The coloured segments of the trajectory represent the path lengths of the identified reversals. In the right panel, the black line connects the endpoints of the observation gap. The median scatter position over the whole time range is marked with a red plus sign, while the dashed red line represents the central jet axis, connecting the median QSC position to the radio core.}
\label{fig:trajectory_smooth_clean}
\end{figure*}

\paragraph{Smoothing and refinement of the QSC trajectory.} The trajectory smoothed by moving average procedure on timescales of 1.5~yr is presented in Fig.~\ref{fig:trajectory_smooth}. The number of averaged positions is reduced to 139 due to boundary handling at the epoch endpoints and a gap in observations. Each averaged position is associated with averaged asymmetric position error, which is shown as an ellipse, with its major axis oriented towards the core. The asymmetric $1\sigma$ position errors are represented by half the lengths of the major and minor axes. The mean position errors are relatively small and generally do not overlap, except in regions where the QSC positions exhibit clustering. Following Arshakian et al. (2024), we consider displacement measurement to be unreliable if the error ellipses of two consecutive positions intersect along the displacement vector. We adopted their algorithm based on this criterion and applied it to the averaged positions. For clarity, we present the refined QSC trajectory in two time intervals (Fig.~\ref{fig:trajectory_smooth_clean}). This procedure further reduces the number of averaged positions by 59\%, resulting in a total of 68.

The overall refined trajectory consists of quasi-linear or arc-shaped patterns followed by loop-like patterns or sharp turns, closely resembling the motion of QSC C7 in the BL~Lac jet \citep{arshakian24}. To determine whether the refined trajectory deviates from randomness, we analyse the angles between successive displacement vectors. In the case of purely random motion, the ratio of the number of angles greater than $90^\circ$ to those less than $90^\circ$ should be approximately unity; that is, $n(>90^\circ)/n(<90^\circ) \approx 1$. Our estimate yields a ratio of $42/13 \approx 3.2$, indicating that forward-directed displacement vectors outnumber backward-directed ones by a factor of 3.2. This evidence indicates that the trajectory is not random but follows a directed path with subsequent reversals.

\paragraph{Characteristics of reversals.} We apply the formalism described in \cite{arshakian24} to characterize the reverse motion of the QSC. We classify the trajectory as reversible (R-type) if the turning angle is less than $90\degr$ and combo reversal (RC-type) if a reversal consists of multiple sub-reversals (Table~\ref{table:char_rev_combo_rev}), for example, two pivot points are encompassed within the loop-like trajectory spanning 1998.41$-$2000.44 (Fig.~\ref{fig:trajectory_smooth_clean}, yellow line, left panel).

We define $t_{\rm R}$ as the epoch of the pivot point of the reverse, while $t_{\rm s}$ and $t_{\rm e}$ represent the estimated epochs of the start and end of the reverse motion, respectively. The duration of the reversal is given by $\Delta t_{\rm R} = t_{\rm e} - t_{\rm s}$, and $\Delta l_{\rm R}$ denotes the length of the reversal.

We identified 12 reversals, 10 R-type and one RC-type (which consists of two R-type), over a 27-year period, with pivot points marked by red circles in Fig.~\ref{fig:trajectory_smooth_clean} (see Table~\ref{table:char_rev_combo_rev}). This yields an estimated reversal frequency of $f_{\rm R} = 0.44$ per year. 
Reversals occur over a range of temporal (or spatial) scales, ranging from about $0.04$~ mas to $0.12$~ mas. Reversals on small spatial scales are likely to be loop-like or arc-like but appear R-type because of smoothing.

Most R-type reversals exhibit sharp turns, except for two cases where the turning angles approach $90\degr$ (see the right panel, red and orange lines). In these instances, the QSC likely follows a loop-like trajectory that appears smoothed due to the 1.5-year averaging applied to the trajectory. As a result, reversal characteristics such as $f_{\rm R}$ and $\Delta l_{\rm R}$ should be considered as lower and upper limits, respectively.

The direction of reversal at the pivot point is clockwise (CW) in 8 cases and counter-clockwise (CCW) in 3 cases (Table~\ref{table:char_rev_combo_rev}). Notably, eight consecutive reversals (numbered 3 to 10) follow a CW direction over a 15.52-year period (excluding a 2.36-year observation gap). The probability of this occurring by chance is very small, only 0.4\%.

We analysed the on-sky distribution of pivot points, specifically examining their radial distances from the median centre and their azimuthal angles relative to the jet axis. The azimuthal angle distribution of the pivot points appears uniform, while most radial distances are concentrated within 0.1~mas from the median centre. No significant differences are found in these distributions between CW and CCW motion.

\paragraph{Estimates of characteristics of transverse waves.} 
\cite{arshakian24} demonstrate that, in the seagull-on-wave model, the reversing trajectories of the QSC are the on-sky projections of its motion, induced by the passage of a relativistic transverse wave. In this scenario, the amplitude and frequency (or wavelength) of the wave approximately correspond to the frequency and half-length of the reversal, with $f_{\rm w} = f_{\rm R} \sim 0.5$~yr$^{-1}$, $A_{\rm w} \approx \Delta l_{\rm R} /2 \sim 0.06$~mas and $\sigma_{\rm A_{\rm w}} = 0.03$~mas. The speeds with which the QSC passes the reversal patterns are superluminal (Table~\ref{table:char_rev_combo_rev}). The superluminal speeds can also be explained within the framework of the seagull-on-wave model, where the QSC moves at a subluminal speed in the host galaxy's rest frame, but appears superluminal in the observer's frame due to the relativistic motion of the transverse wave. In reality, these speeds represent lower limits, as the QSC trajectory is shortened due to smoothing effects.

The propagation speed of the transverse wave can be estimated as $\beta_{\rm w} = \lambda_{\rm w} f_{\rm w}$, where $\lambda_{\rm w} = \lambda_{\rm w, app} / \sin \theta$ is the wavelength and $\lambda_{\rm w, app}$ is the apparent wavelength. Measurement of the apparent wavelength is challenging due to the non-periodic and swirling nature of the transverse waves, compounded by the small angle between the wave propagation direction and the line of sight.
\\

Finally, we repeated the analysis of this section for a smoothing time window of 2~yr. The coefficient characterizing the randomness of motion increases to 2.1, further supporting the directional motion of the QSC. Additionally, the number of reversals decreases to 10, predominantly preserving those occurring on larger scales.

\begin{table*}[htbp]
\caption{Characteristics of reversals and reversal combinations.}
\label{table:char_rev_combo_rev}
\centering
\begin{tabular}{cccccccccc}
\hline
\multicolumn{1}{c}{} & \multicolumn{1}{c}{Type} & \multicolumn{1}{c}{Epoch} & \multicolumn{1}{c}{Start} & \multicolumn{1}{c}{End} & \multicolumn{1}{c}{Time interval} & \multicolumn{1}{c}{Length} & \multicolumn{1}{c}{Speed} & \multicolumn{1}{c}{Spatial scale} & \multicolumn{1}{c}{cw/}\\
\multicolumn{1}{c}{} & \multicolumn{1}{c}{} & \multicolumn{1}{c}{(yr)} & \multicolumn{1}{c}{(yr)} & \multicolumn{1}{c}{(yr)} & \multicolumn{1}{c}{(yr)} & \multicolumn{1}{c}{(mas)} & \multicolumn{1}{c}{(c)} & \multicolumn{1}{c}{(mas)} & \multicolumn{1}{c}{ccw} \\
\hline
1 & R$^{a}$ & 1996.78 & 1996.37 & 1998.41 & 2.03 $(\times2)$ & 0.14 $(\times2)$ & 4.37 & 0.12 & ccw \\
2 & RC$^{b}$ & 1999.26 & 1998.41 & 2000.44 & 2.03 & 0.09 & 2.89 & 0.04 & ccw \\
3 & R & 2001.61 & 2000.44 & 2003.74 & 3.29 & 0.22 & 4.22 & 0.14 & cw \\
4 & R & 2005.20 & 2003.74 & 2005.71 & 1.97 & 0.13 & 4.27 & 0.06 & cw \\
5 & R & 2006.68 & 2005.71 & 2007.17 & 1.46 & 0.12 & 5.27 & 0.10 & cw \\
6 & R & 2008.48 & 2007.17 & 2009.09 & 1.92 & 0.08 & 2.75 & 0.05 & cw \\
7 & R & 2009.66 & 2009.09 & 2010.07 & 0.98 & 0.08 & 5.24 & 0.05 & cw \\
8 & R & 2015.05 & 2014.12 & 2017.31 & 3.18 & 0.17 & 3.47 & 0.10 & cw \\
9 & R & 2017.65 & 2017.31 & 2018.31 & 1.00 & 0.08 & 4.98 & 0.05 & cw \\
10 & R & 2019.49 & 2018.31 & 2019.64 & 1.34 & 0.12 & 5.76 & 0.07 & cw \\
11 & R & 2020.16 & 2019.64 & 2021.14 & 1.50 & 0.11 & 4.79 & 0.08 & ccw \\
\hline
Mean$^{c}$: & $-$ & $-$ & $-$ & $-$ & $1.87 \pm 0.26$ & $0.12 \pm 0.01$ & $4.36 \pm 0.33$ & $0.07 \pm 0.01$ & $-$ \\
\hline
\end{tabular}
\tablefoot{The columns for reversals are as follows: reversal type; the epochs of the beginning and end of the reversal, $t_{\rm s}$ and $t_{\rm e}$; reversal time interval $\Delta t_{\rm R} = t_{\rm e} - t_{\rm s}$; length of the reversal $\Delta l_{\rm R}$; speed of the QSC component $\beta_{\rm s}$; spatial scale, representing the largest extent of the reversal; and the direction of the reverse motion. \\
\tablefoottext{a}{Only part of the reversal is observed. The length and time interval of the reversal are taken to be equal to twice the measured length and time interval.} \\
\tablefoottext{b}{This is a combo reversal, encompassing two smaller-scale reversals occurring at epochs 1999.26 and 2000.07, both exhibiting CCW motion (yellow line in Fig.~\ref{fig:trajectory_smooth_clean}, left panel). The pivot point of the combo reversal is identified at epoch 1999.26.}  \\
\tablefoottext{c}{Mean values are calculated excluding the characteristics of the first reversal, as only a portion of it is observed.}
}
\end{table*}

\section{Discussions}
\label{sec:discussions}
BL~Lac objects and FSRQs differ significantly in several key characteristics, including optical spectral properties, jet power, large-scale radio morphology (FR~I vs. FR~II), spectral energy distributions (synchrotron-dominated vs. inverse Compton-dominated), and the relative strength of their $\gamma$-ray emission. Here, we compare the properties of relativistic transverse waves observed in two blazars: BL~Lac \citep{cohen15, arshakian24} and the FSRQ 3C~279, as analysed in this work. The wave frequency and mean wave amplitude are estimated to be $0.5$~yr$^{-1}$ and 0.06~mas (0.38~pc) for the 3C~279 and $0.8$~yr$^{-1}$ and 0.028~mas (0.037~pc) for the BL~Lac. While the wave frequencies are comparable, the mean wave amplitude of the 3C~279 is larger than that of the BL~Lac by a factor of ten. Assuming that the transverse wave is an Alfvén wave, and the transverse magnetic energy $B_{\perp}$ dominates over the transverse kinetic energy of the jet plasma, the total energy of the wave in the frame of the jet $E_{\rm tot} \propto B_{\perp}^2 V \approx (A_{\rm w}f_{\rm w} B_0)^2 V$, where the magnetic energy of the perturbation $B_{\perp} \approx A_{\rm w} f_{\rm w} B_0$, $B_0$ is the mean large-scale magnetic field that exists before perturbations by Alfvén waves, and $V$ is the total volume of the jet segment containing the wave. Assuming that the values of $B_0$ and $V$ are approximately equal for BL~Lac and 3C~279, we obtain the ratio of their total energies $(0.5 \times 0.38)^2/ (0.8 \times 0.037)^2 \approx 40$; that is, the total transverse wave energy of 3C~279 is approximately $10^{1.6}$ times higher than that of the BL~Lac wave. This difference is primarily due to the significantly larger mean wave amplitude of the quasar 3C~279.

The mean speeds of the QSCs in 3C~279 and BL~Lac are superluminal, $15c$ and $2.5c$ \citep{arshakian24}, with the former being about six times higher. This difference can be  attributed to a smaller viewing angle of the jet, a higher inclination angle and speed of the transverse wave in 3C~279, or a combination of these factors (see Eq.~\ref{eq:beta_s_sagittal}).

Another significant difference lies in the jet structure on pc-scales: the deprojected distance of the QSC from the core in 3C~279 ($\approx 125$~pc) is approximately 40 times greater than that of the QSC in BL~Lac ($\approx 3$~pc), assuming a jet viewing angle of $1\degr$ adopted in this work and $6\degr$ from \cite{arshakian24}, respectively. The QSC in the BL~Lac jet is believed to be a recollimation shock at 15~GHz \citep{cohen14}, supported by high-resolution observations at 22~GHz with \textit{RadioAstron} and 43~GHz with VLBA \citep{gomez16}. Relativistic magnetohydrodynamics simulations in \cite{gomez16} show that recollimation shocks appear at relatively short distances from the core due to the jet's high speed and a decreasing pressure gradient in the ambient medium, possibly caused by interactions with the broad-line region ($0.1-1$~pc) or torus ($\lesssim 10$~pc). On the other hand, the location of the QSC in 3C~279 at a distance of about 125~pc from the core is beyond the broad-line region and torus, indicating that the recollimation shock is generated by interaction with a hot gas region.

\cite{burd22} show that the transition of the jet geometry from parabolic to conical in 3C~279 occurs at a separation of 0.6~mas (deprojected distance of $\approx 210$~pc), likely marking the transition from a magnetically dominated to a particle-dominated jet \citep{kovalev20}. In BL~Lac, the break point of the jet geometry transition occurs at a separation of 2.5~mas \citep[deprojected distance of $\approx 30$~pc;][]{kovalev20}, well beyond the torus region. This supports the dominance of helical magnetic energy over kinetic energy of the jet at large distances from the central black hole, and consequently the presence of relativistic Alfvén waves extending up to the jet geometry transition location.

Lastly, the ratio of rms$_{\rm c}$/rms$_{\rm s}$ is $\approx 1.6$ for 3C~279 and $\approx 1.3$ for BL~Lac \citep{arshakian20}, indicating that the variation in the displacements of the core is more pronounced in 3C~279. This could be attributed to stronger opacity changes (or higher particle density variations) during radio flares or a stronger magnetic field in the core region, as well as geometric effects such as wobbling of the core \citep[see a detailed discussion in][]{arshakian20}.

\section{Summary}
\label{sec:summary}
We have investigated the dynamics of the QSC in the FSRQ 3C~279 using 15~GHz VLBA monitoring data from the MOJAVE programme. The QSC is identified in 161 epochs spanning over 27 years, with its median position located at a separation of $\sim 0.35$~mas from the radio core (or projected distance of $\approx 2.2$~pc).

Studies of the on-sky distribution of QSC positions have proven to be a valuable tool for analysing changes in the jet direction and cone opening angle. Our analysis reveals that the jet axis has changed direction by approximately $20\degr$ over this 27-year period. During this time, the jet axis has gradually shifted southwards, showing signs of nutation. The apparent cone angle of the QSC is in good agreement with the apparent jet cone angle inferred from the jet width and the motion of components downstream of the QSC and it varies over time, with a mean value of about $16\degr \pm 2\degr$, corresponding to a deprojected intrinsic cone angle of $0.3\degr \pm 0.03\degr$, assuming a jet viewing angle of $1\degr$. The scatter of QSC positions across the jet axis reflects the maximum amplitude of the transverse waves propagating through the QSC.

The on-sky distribution of the QSC's apparent displacement vectors reveals an excess in both number and magnitude along the jet axis, suggesting strong anisotropic displacements of the core in the jet direction. To quantify this effect, we applied a method that disentangles apparent displacements and allows for statistical estimation of the intrinsic apparent displacements of both the QSC and the core. We found that the intrinsic rms$_{\rm c}$ of core displacements exceeds the intrinsic rms$_{\rm s}$ of the QSC by a factor of $\approx 1.6$. Using the estimated mean intrinsic displacement of the QSC and the standard deviation of its displacements, we estimated the mean speed of the QSC as $\overline{\beta}_{\rm s} \approx 10 \pm 1.3$. Within the seagull-on-wave model, the observed superluminal speed of the QSC provides evidence of relativistic transverse waves propagating through it. We further estimated the approximate mean speed of the transverse wave ($\Gamma_{\rm w} \approx 60$) and its mean inclination angle ($\overline{\varphi} \approx 0.2\degr$) relative to the jet axis. 

The characteristics of the QSC’s reversing trajectory and its classification enable the amplitudes and frequencies of these relativistic transverse waves to be determined. By analysing the smoothed and refined trajectory of the QSC, we estimated a wave frequency of $\sim 0.5$~yr$^{-1}$ and a mean wave amplitude of $\overline{A}_{\rm w} \approx 0.06 \pm 0.005$~mas ($\approx 0.38$~pc), which in fact represent the lower and upper limits, respectively. The estimate of $\overline{A}_{\rm w}$ is consistent with the mean maximum wave amplitude derived from the transverse scatter of QSC positions, $\overline{A}_{\rm w} < \overline{A}_{\rm w, max} = 0.1$~mas.  

We compared the characteristics of the QSC and transverse waves in the quasar 3C~279 and the BL~Lac object. In 3C~279, the magnetic energy of the transverse wave, its mean wave amplitude, the mean speed of the QSC, and its deprojected distance from the core are significantly larger than those in BL~Lac. However, the wave frequency and the distance to the break points marking the transition of the jet shape from parabolic to conical are comparable in both sources. The jets of 3C~279 and BL~Lac are believed to host strong helical magnetic fields at distances of several tens of parsecs from the central black hole, supporting the presence of relativistic transverse waves beyond the torus region.

We conclude that studying the dynamics of the QSC is a valuable tool for probing the presence of relativistic transverse waves in the jets of BL Lac objects and FSRQs. It also enables the determination of their key characteristics, including frequency, speed, amplitude, and magnetic energy.

\begin{acknowledgements}
We thank the MOJAVE team for comments on the manuscript and the anonymous referee for careful reading and valuable suggestions. The VLBA is a facility of the National Radio Astronomy Observatory, a facility of the National Science Foundation that is operated under cooperative agreement with Associated Universities, Inc. This research has made use of data from the MOJAVE database that is maintained by the MOJAVE team \citep{lister18}. The research was supported by the YSU, in the frames of the internal grant. 
\end{acknowledgements}

\bibliographystyle{aa} 
\bibliography{refs_tga}

\end{document}